# Translating Machine Learning Interpretability into Clinical Insights for ICU Mortality Prediction

Ling Liao[1,2*], Eva Aagaard[3*]

1. Biomedical Deep Learning LLC, MO, U.S.A, 63130
2. Computational and Systems Biology, Washington University, MO, U.S.A, 63130
3. School of Medicine, Washington University School of Medicine, MO, U.S.A, 63130

* Authors to whom correspondence should be addressed.
Correspondence: Ling Liao (lingliao@wustl.edu), Eva Aagaard (aagaarde@wustl.edu)

**Abstract**
Current research efforts largely focus on employing at most one interpretable method to elucidate machine learning (ML) model performance. However, significant barriers remain in translating these interpretability techniques into actionable insights for clinicians, notably due to complexities such as variability across clinical settings and the Rashomon effect. In this study, we developed and rigorously evaluated two ML models along with interpretation mechanisms, utilizing data from 131,051 ICU admissions across 208 hospitals in the United States, sourced from the eICU Collaborative Research Database. We examined two datasets: one with imputed missing values (130,810 patients, 5.58% ICU mortality) and another excluding patients with missing data (5,661 patients, 23.65% ICU mortality). The random forest (RF) model demonstrated an AUROC of 0.912 with the first dataset and 0.839 with the second dataset, while the XGBoost model achieved an AUROC of 0.924 with the first dataset and 0.834 with the second dataset. Consistently identified predictors of ICU mortality across datasets, cross-validation folds, models, and explanation mechanisms included lactate levels, arterial pH, body temperature, and others. By aligning with routinely collected clinical variables, this study aims to enhance ML model interpretability for clinical use, promote greater understanding and adoption among clinicians, and ultimately contribute to improved patient outcomes.

**Introduction**
Trustworthy ML models are essential for predicting high-stakes clinical outcomes.[1,2,3,4,5] However, their adoption has been hindered by their propensity to generate biased, incorrect and occasionally harmful results.[6,7,8,9,10,11] Current strategies are moving towards 1) using multi-institutional data for model development, 2) applying some level of explanation to interpret how the predictions have been made, and 3) involving clinical knowledge evaluation in the loop.[2,8,12,13] Despite these efforts, significant challenges remain: 1) multi-institutional data may not adequately account for the substantial variations in admission criteria, patient populations, and baseline physiological states across different hospitals , and 2) while several articles address model interpretability, most rely on a single method, like SHAP. Together, these limitations can produce varying interpretations due to differing assumptions and algorithms.

In this study, we developed two sets of ML models and explanation mechanisms to reveal key clinical features and clusters influencing the models' ability to predict ICU mortality within the first 24 hours after admission. By comparing the results with interpretations and validating them using clinical knowledge, we aimed to bolster confidence in the models' applicability for clinical trials and their potential to drive clinical improvements.

---

* The differences between 130,810 and 131,051 is because we found 214 records with some missing feature values.

**Methods**
**Study design and setting** We developed two sets of ML models, RF and XGBoost, alongside explanation mechanisms to identify key clinical features and clusters influencing the models' ability to predict ICU mortality within the first 24 hours of admission. The study analyzed 131,051 ICU admissions across 208 U.S. hospitals from 2014 to 2015, using a subset of data from eICU-CRD.[3,14,] Patients under 16 years of age, those with admissions shorter than 6 hours, and those with prior ICU stays were excluded.[3] Critical missing variables were either imputed or the corresponding patients were excluded from model development. Five-fold cross-validation was employed to assess model performance. The AUROC and AUPRC were used to evaluate the models' performance. Built-in feature importance from RF, SHAP values, and K-Means clustering were utilized to determine the significance of various features and clusters contributing to the models' predictions.

**Data preprocessing** After the removal of data from 241 patients with remaining missing values, two strategies were employed to the collected dataset. The first strategy, referred to as Dataset 1, involved utilizing the complete dataset comprising 130,810 patients with a 5.58% ICU mortality rate and imputing missing variables. The second strategy, identified as Dataset 2, entailed using data without imputation by excluding all patient records that contained any missing values. As a result, Dataset 2 included 5,661 patients, which had a significantly higher ICU mortality rate of 23.65%. These contrasting approaches aimed to assess the effects of data completeness and the impact of imputation methods on model performance, thereby providing insights into the robustness and applicability of the models under varying data quality conditions. The potential influence of hospital variability on patterns of missing data and model performance is discussed in detail in the discussion section.

**Imputation methods** The imputation method used to address missing values in Dataset 1 was detailed by Raffa et al. (2022).[3] Briefly, patients with complete data for given variables were leveraged to calculate or predict values for cases with missing data using four algorithms. Algorithm 0 involved replacing missing numeric values with the median and categorical values with the mode from complete cases. Algorithm 1 developed imputation models using XGBoost to predict values for missing variables. Algorithm 2 similarly used XGBoost but excluded variables of the same type (e.g., not using minimum body temperature to predict maximum body temperature). Algorithm 3 combined the strategies from strategy 1 and 2 by applying strategy 2 to patients missing similar variables and strategy 1 to others. For each variable with missing values, the final algorithms were selected using nested cross-validation, optimizing for minimizing mean squared error for numeric variables and maximizing accuracy for categorical variables.

**Feature exclusion and modifications** Features excluded from both Dataset 1 and Dataset 2 include 'patientunitstayid', 'encounter_id', 'hospital_death', and 'partition'. Hospital death was removed from the input features due to its significant overlap with ICU mortality to avoid leakage between input features and output mortality. In Dataset 2, 'dcs_group' (a simplified version of the Glasgow Coma Scale (GCS) that categorizes patients into four groups based on their eye, motor, and other responses) and 'group' (indicating the diagnosed disease, such as sepsis, etc.) were not available. Instead, 'gcs_eyes_apache,' 'gcs_motor_apache,' 'gcs_unable_apache,' and 'gcs_verbal_apache' were incorporated.

The feature names have been changed as follows: 'vent', 'dx_class', and 'dx_sub' from Dataset 1 correspond to 'ventilated_apache', 'apache_2_diagnosis', and 'apache_3j_diagnosis' in Dataset 2. Notably, 'group' is now considered equivalent to 'apache_3j_diagnosis'.

For further feature modifications, in Dataset 1, 26 features that end with '_avg' or '_diff' (13 each) have been replaced by features ending with '_max' and '_min', with different values derived from the real patient data. Post-processing, Dataset 1 includes 66 input features, while Dataset 2 includes 68, with 'icu_death' as the label for both. The detailed descriptions of each feature are comprehensively documented in Raffa et al. (2022).[3] Upon reviewing the source article, we identified a necessary correction to the feature description of 'd1_albumin_max'. Initially, it was described as the lowest albumin concentration in the patient's serum during the first 24 hours of their unit stay. We have corrected this to reflect the highest albumin concentration of the patient in their serum during the first 24 hours of their unit stay.

**Model development** RF and XGBoost were selected for model development due to their widely favored ability for handling large datasets, modeling complex relationships, and delivering high-accuracy predictions in various fields such as healthcare, education, and finance involving tabular and feature-rich data.[1,2,11,12,15,16,17,18] The input data, either Dataset 1 or Dataset 2, was randomly shuffled first, then split into 5 equal parts, or folds. The model was trained on four of the folds and tested on the remaining fold. This process was repeated five times, with each fold being used as the test set exactly once. Within each fold, the performance of the model was evaluated using AUROC and AUPRC.

**Model interpretation** Feature importance refers to the individual contribution of each feature to the predictive ability of the models, aiding in understanding which features most significantly affect their performance.[19] For RF model, the built-in feature importance was calculated to explain the model's prediction based on total decrease in node (individual decision point within a tree) impurity.[11,18] Each time a feature is used to split a node, the algorithm calculates the reduction in impurity resulting from the split and attributes this reduction to this feature. The decreases are accumulated across all trees and normalized to produce a global importance score for each feature. For a given feature $j$, the feature importance $\aleph_j$ is computed as:

$$\aleph_j = \frac{1}{T}\sum_{t=1}^{T} \sum_{n \in Nodes_j^{(t)}} \left(\frac{N_n}{N} \cdot \Delta i_n\right) \quad (1)$$

Where $T$ is the total number of trees in the forest, $Nodes_j^{(t)}$ denotes the set of nodes in tree $t$ where feature $j$ is used for splitting, $N_n$ is the number of samples reaching node $n$, N is the total number of training samples, $\Delta i_n$ is the decrease in impurity at node $n$, calculated as:

$$\Delta i_n = i_{parent} - \left(\frac{N_{left}}{N_{parent}} \cdot i_{left} + \frac{N_{right}}{N_{parent}} \cdot i_{right}\right) \quad (2)$$

Where $i = 1 - \sum p_k^2$ is the Gini impurity, with $p_k$ the proportion of class $k$ samples at the node. $i_{parent}$, $i_{left}$, and $i_{right}$ denote the Gini impurity values of the parent, left child, and right child nodes, $N_{left}$ and $N_{right}$ are the number of samples in the left and right child nodes, $N_{parent} = N_{left} + N_{right}$.

SHAP values were used in conjunction with the XGBoost model. SHAP is a unified framework for model interpretability based on cooperative game theory which assigns each feature a locally accurate and globally consistent importance score by quantifying its marginal contribution to individual model predictions.[19] When applied to XGBoost, the SHAP value feature importance $\emptyset_i$ for feature $i$ is calculated as:

$$\emptyset_i = \sum_{S \subseteq F\{i\}} \frac{|S|!(|F|-|S|-1)!}{|F|!} [f(S \cup \{i\}) - f(s)] \quad (3)$$

Where $F$ is the full set of input features, $S$ is a subset of $F$ not including feature $i$, and $f(S)$ is the model's prediction using only features in subset $S$.

Building upon the concept of feature importance, we introduced *cluster importance* to measure the collective impact of potentially related feature groups (clusters) on the model's outcome using K-Means clustering. K-Means clustering is an unsupervised learning algorithm that partitions a dataset into $K$ distinct, non-overlapping clusters by initializing (K) centroids, assigning data points to the nearest centroid, and iteratively updating the centroids until they stabilize.[20] In our study, we set *K=20* to divide the features into 20 clusters based on their importance scores derived from the trained RF or XGBoost model, aiming to minimize within-cluster variance. Given a set of $n$ features represented by their importance values $\{x_1, x_2, \ldots, x_n\}$, K-Means seeks to solve:

$$\arg min_C \sum_{k=1}^{K} \sum_{x_i \in C_k} ||x_i - \mu_k||^2 \qquad (4)$$

Where $C_k$ is the set of features assigned to cluster $k$, and $\mu_k$ is the centroid of cluster $k$. The algorithm iteratively assigns each feature to the nearest cluster centroid and then updates the centroids until convergence.

**Results**

With RF, Dataset 1 (a positive case rate of 0.0558) yielded an AUROC of 0.912±0.002 and an AUPRC of 0.484±0.007, whereas Dataset 2 (a 0.2365 positive case rate) resulted in an AUROC of 0.839±0.010 and an AUPRC of 0.637±0.029. Comparatively, the XGBoost model achieved an AUROC of 0.924±0.003 and an AUPRC of 0.549±0.007 with Dataset 1, and an AUROC of 0.834±0.011 and an AUPRC of 0.648±0.023 with Dataset 2.

Figure 1. Performance of RF and XGBoost Models on Dataset 2. (A) AUROC curves for the RF model. (B) AUPRC curves for the RF model. (C) AUROC curves for the XGBoost model. (D) AUPRC curves for the XGBoost model. AUROC: Area Under the Curve - Receiver Operating Characteristic. AUPRC: Area Under the Curve - Precision-Recall Curve. Baseline indicates the statistical positive rate, a 23.65% ICU mortality rate in Dataset 2.

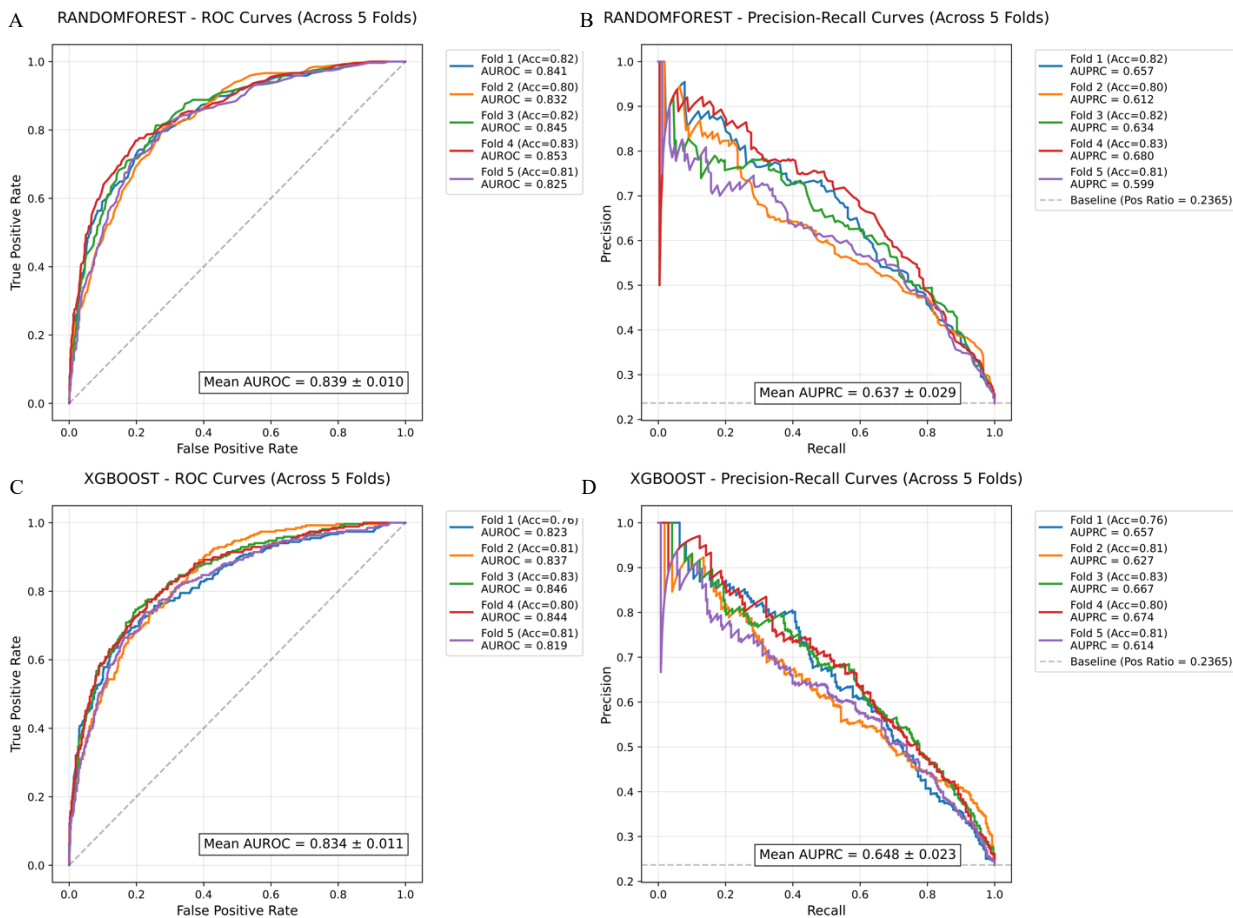

Fig. 1 details the performance of the two models applied to Dataset 2. In panels A and C, y-axis represents the true positive rate, while x-axis denotes the false positive rate. The colored lines display AUROC curves for each fold within the dataset. Panels B and D illustrate Precision-Recall curves for each fold, where y-axis represents precision and x-axis denotes recall. The baseline reflects the natural prevalence of positive cases, corresponding to the dataset's ICU mortality rate of 0.2365. The models demonstrated their ability to distinguish patients at risk of ICU mortality with AUPRC values of 0.637 and 0.648, respectively, achieving approximately 2.7 times better performance than the baseline. Additionally, the results demonstrate that the two models exhibit relatively consistent AUROC and AUPRC values across all five folds and both datasets. However, as shown in Supplementary Fig. 1, both models achieved higher AUROC values but lower AUPRC values on Dataset 1 than on Dataset 2.

Fig. 2 depicts the feature and cluster importance heatmaps for the RF and XGBoost models applied to Dataset 2. On the y-axis, features are listed with the highest importance at the top and the lowest importance at the bottom, while the x-axis shows the unique clusters, numbered 1 to 20, across folds 1 to 5. Above the heatmap, clusters are ranked by their aggregated importance from most important (cluster 1 for each fold) to least important (cluster 20 for each fold). Note that the cluster numbers on the x-axis correspond to the same clusters ranked by their aggregated importance. Specifically, Fig. 2A presents the feature and cluster importance for the RF model, whereas Fig. 2B illustrates the importance map for the XGBoost model.

The averaged top 10 important features across 5-folds in the RF model are provided in Table 1. Collectively, these features provide a robust framework for evaluating the severity of a patient's condition and the risk of mortality.

Table 1. Top 10 Important Features from RF Model (Averaged across 5-folds)

| Feature | Category | Function |
| --- | --- | --- |
| d1_lactate_max | Lactate levels | Metabolic status and potential shock |
| d1_lactate_min | Lactate levels | Metabolic status and potential shock |
| d1_arterial_ph_max | Arterial pH | Acid-base balance |
| apache_3j_diagnosis | APACHE III-J diagnosis | Primary reason for ICU admission |
| d1_inr_max | INR | Coagulation status |
| d1_spo2_min | Peripheral oxygen saturation | Respiratory function and oxygen therapy efficiency |
| gcs_motor_apache | Glasgow Coma Scale (motor) | Neurological status |
| d1_pao2fio2ratio_min | Fraction of inspired oxygen | Respiratory function and oxygen therapy efficiency |
| d1_hco3_min | Bicarbonate levels | Acid-base balance |
| d1_creatinine_max | Creatinine levels | Renal function |

For XGBoost, eight out of the top 10 features are the same as those in RF. The differing features in XGBoost are d1_temp_min and gcs_eyes_apache, while RF includes d1_hco3_min and d1_creatinine_max. Core temperature (d1_temp_min) is crucial for detecting infections and metabolic imbalances, directly affecting patient stability. The eye-opening component of the Glasgow Coma Scale (gcs_eyes_apache) provides additional insights into neurological status and responsiveness.

In the cluster-level interpretation, we evaluated whether features of low importance might cluster with highly important features to more substantially contribute to the model's predictive ability. The results indicate that, in both Fig. 2A and Fig. 2B, the top 10 important features are mostly distinct or tend to cluster with neighboring highly important features rather than significantly less important ones, collectively contributing to the model's predictive capacity. The feature and cluster importance maps of the models using Dataset 1, as detailed in Supplementary Fig. 2, reveal similar trends.

Another noteworthy finding from Fig. 2 and Supplementary Fig. 2 is that many individually less important features, when clustered together, have a considerable impact on the models' predictions across both datasets. For example, d1_mbp_min, which represents the patient's lowest mean blood pressure during the first 24 hours of their unit stay, is not among the top 10 important features in both Fig. 2A and Fig. 2B. However, when clustered with other features such as d1_pao2fio2ratio_max (the highest fraction of inspired oxygen for the patient during the first 24 hours), d1_inr_min (the lowest international normalized ratio for the patient during the first 24 hours), d1_sysbp_min (the patient's lowest systolic blood pressure during the first 24 hours), and other features, this cluster's importance ranks within the top 10 primary clusters contributing to the model's predictive ability in Fig. 2A, fold 1. When clustered with other below top-10 features, while those features can vary across folds, clusters including d1_mbp_min are listed in the top 10 clusters across all five folds in both Figures. 2A and 2B.

Figure 2. Feature and Cluster Importance Map of the Two Models Using Dataset 2 (23.65% ICU Mortality Rate). (A) Importance map for the RF model. (B) Importance map for the XGBoost model.

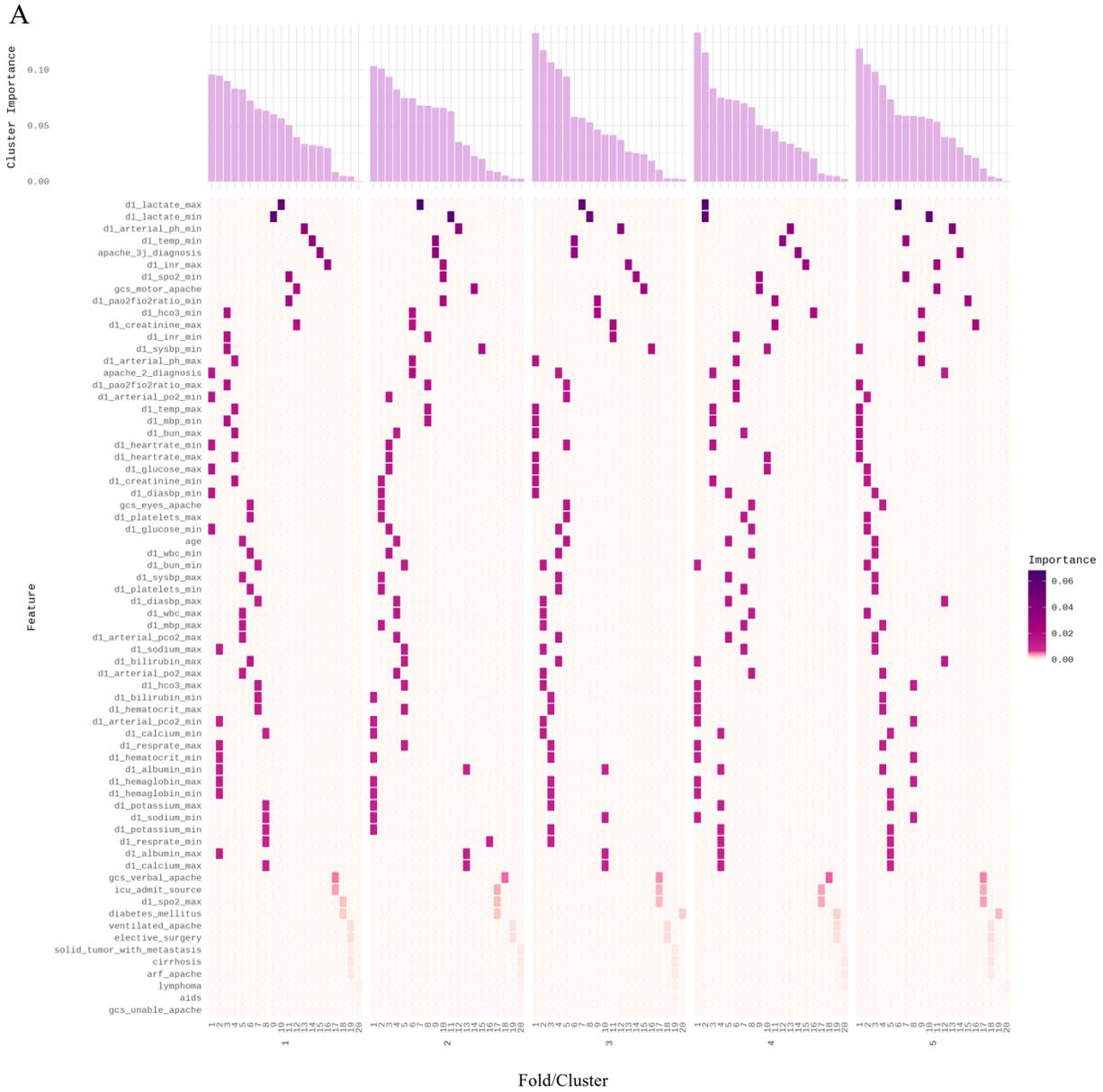

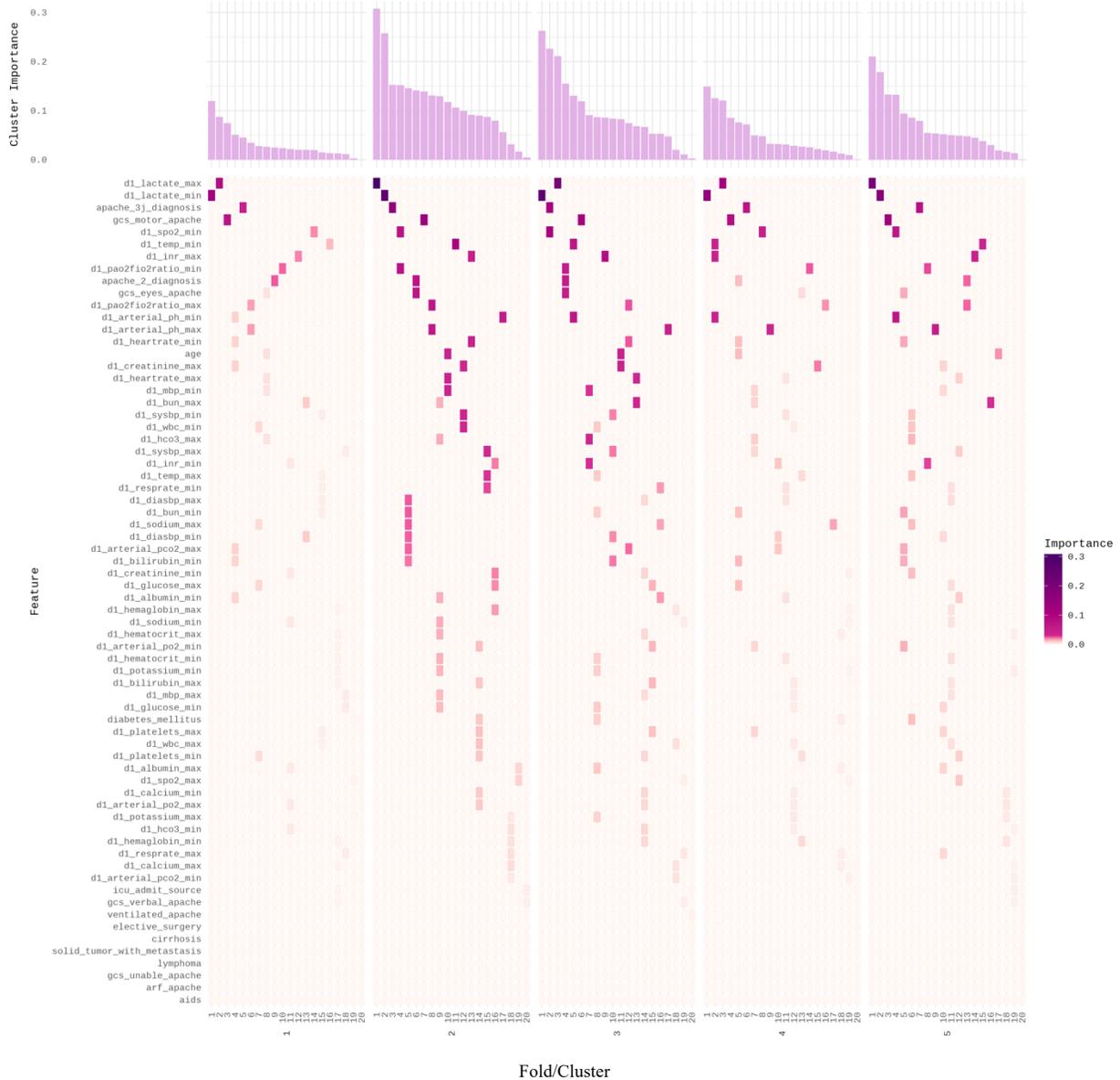

## Discussion

In this interpretability-focused study, we developed two sets of ML models along with interpretation mechanisms to pinpoint key clinical features and clusters that impact the models' ability to predict ICU mortality within the first 24 hours of admission. Both the RF and XGBoost models achieved comparable performance metrics across two datasets.

In Dataset 1, RF achieved an AUROC of 0.912 and an AUPRC of 0.484, while XGBoost reached an AUROC of 0.924 and an AUPRC of 0.549. For Dataset 2, RF recorded an AUROC of 0.839 and an AUPRC of 0.637, whereas XGBoost attained an AUROC of 0.834 and an AUPRC of 0.648. The differences across datasets may be attributed to multiple factors including the larger size of Dataset 1, clinical variability among different hospitals and their associated patterns of missing data, the imputation methods employed, and the higher prevalence of positive cases in Dataset 2.

The general performance of AUROC and AUPRC curves across all five folds using data spanning 208 U.S. hospitals, as well as the identified primary features consistent with established predictors of ICU mortality[21,22,23,24], supports the potential applicability of these models in diverse clinical environments. This conclusion is drawn from: 1) the robustness demonstrated by the models; 2) the consistent metrics across folds suggesting reliable predictions and resistance to overfitting; 3) the use of five-fold cross-validation, wherein the datasets are randomly shuffled and divided into five equal parts, with each part used once as the test set while the remaining four parts are used for training. This approach, incorporating testing data from numerous hospitals in a diverse dataset, showcases the model's ability to generalize across varied clinical settings; and 4) the consistently identified primary features further strengthens this assertion.

Comparatively, the GOSSIS model developed by Raffa et al. (2022) achieved an AUROC of 0.904 on Dataset 1, which included an additional 241 patient records.[3] The APACHE-IVa model achieved an AUROC of 0.869 on the eICU-CRD dataset, where patients with missing data on non-Acute Physiology Score III components were excluded, and missing values were imputed using an undisclosed method.[3] Notably, AUPRC values were not reported for either the GOSSIS or APACHE-IVa models, nor were AUROC and AUPRC curves or interpretation feature importance provided, limiting their transparency and clinical applicability.

Although strategies for developing ML models to predict high-stakes clinical outcomes are indeed moving towards incorporating multi-institutional data, model interpretation, and clinician evaluation,[2,8,12,13,25,26,27] many of these strategies acknowledge limitations regarding external or additional independent validation of their models. In a recent study by Friedman et al. (2025), a multimodal ML model was developed and implemented in real-world clinical settings to predict delirium risk among hospitalized patients.[1] Despite encouraging results, the authors acknowledged limitations in the model's generalizability, such as the lack of external validation and reliance on a unique delirium treatment method.[1] Furthermore, few strategies have recognized the variance of potential conflicting interpretations across different algorithms and explanation mechanisms.

A major strength in our study is the approach to interpretability, utilizing built-in feature importance values from RF, SHAP values, and K-Means clustering. The cross-method consistency in identifying key features, like lactate levels and arterial pH, underscores their role as universally strong indicators of ICU mortality, transcending model-specific biases. This convergence of data-driven insights with clinical intuition not only validates the models' reliability but also shifts the focus from predictive performance to mechanistic understanding—revealing how these factors collectively shape outcomes. By aligning with routinely collected clinical variables, our findings mitigate the gap between interpretable ML and real-world decision-making, empowering clinicians to confidently apply these insights using routine clinical data.

Several limitations should be acknowledged. First, its retrospective design and reliance on publicly sourced data warrant caution regarding the accuracy and representativeness of records from previous visits in electronic health records from numerous hospitals. The potential variability and inconsistencies in documentation practices across institutions could impact the reliability of our findings. Real-time validation in ongoing clinical settings would be necessary to further verify the models' efficacy and generalizability. Secondly, while our study includes a broad dataset, the exclusion of certain patient groups—such as those under 16 years of age, those with admissions shorter than 6 hours, and individuals with prior ICU stays—limit the generalizability of our results to these specific populations.

In conclusion, this interpretability-focused study demonstrates that ML models, specifically RF and XGBoost, can effectively predict ICU mortality within the first 24 hours after admission. Key predictors identified align well with established ICU mortality indicators. The in-general consistent performance and primary features identified across all five folds in datasets spanning 208 U.S. hospitals support the potential applicability of these models in diverse clinical settings in this country. This study aims to enhance clinicians' understanding and adoption of ML models in clinical practice, ultimately leading to improved patient outcomes.

**Data availability:** The data utilized in this study is available at https://physionet.org/content/gossis-1-eicu/1.0.0/. Access to this resource is restricted; proper registration, required training, and a signed data usage agreement may be necessary.

**Code availability**: The code developed in this study is available at https://github.com/lingliao/Interpretable_AL

**Acknowledgements** Liao would like to acknowledge the consistent support from McDonnell International Scholars Academy at Washington University in Saint Louis.

**Author contributions** L.L. and E.A. contributed equally to this study. L.L. and E.A. conceptualized and designed the study. L.L. was responsible for the acquisition, analysis, and interpretation of the data. L.L. and E.A. drafted the manuscript. L.L. and E.A. critically reviewed the manuscript for important intellectual content. Liao performed the statistical analysis. Liao and Aagaard supervised the study. All authors reviewed the manuscript.

**Competing interests** Liao is the founder of Biomedical Deep Learning LLC. Aagaard has no conflict of interest to disclosure.